\pdfoutput=1

\documentclass[11pt]{article}

\usepackage[]{ACL2023}

\usepackage{times}
\usepackage{latexsym}

\usepackage[T1]{fontenc}

\usepackage[utf8]{inputenc}

\usepackage{microtype}

\usepackage{inconsolata}

\usepackage{url}

\usepackage{amsmath}
\usepackage{graphicx}
\usepackage{makecell}
\usepackage{caption}
\usepackage{subcaption}
\usepackage{booktabs}
\usepackage{tabularx}
\usepackage{multirow}
\usepackage{color,soul}
\usepackage{enumitem}
\usepackage{hyperref}
\usepackage{cleveref}

\crefformat{section}{\S#2#1#3} 
\crefformat{subsection}{\S#2#1#3}
\crefformat{subsubsection}{\S#2#1#3}

\usepackage{mathtools}
\usepackage{commath}
\usepackage{placeins}
\usepackage{adjustbox}
\usepackage[bottom]{footmisc}
\usepackage{array, blkarray, makecell}

\usepackage{float}
\usepackage{amssymb}
\usepackage{mhchem}
\usepackage{siunitx}
\usepackage{mathtools, nccmath}

\usepackage{slashbox}
\usepackage{comment}

\usepackage{bm}

\newcolumntype{P}[1]{>{\centering\arraybackslash}p{#1}}

\usepackage[linesnumbered,noline,noend]{algorithm2e}
\usepackage{tablefootnote}
\usepackage{bbm}
\usepackage{xcolor,colortbl}

\usepackage[linesnumbered,noline,noend]{algorithm2e}
\usepackage[colorinlistoftodos]{todonotes}

\newcommand{\pgsname}{HyHTM}

\title{\pgsname: Hyperbolic Geometry based \\ Hierarchical Topic Models}

\author{Simra Shahid \thanks{\enskip Authors contributed equally to the work.} \quad Tanay Anand\footnotemark[1] \quad Nikitha Srikanth\footnotemark[1] \\ {\bf Sumit Bhatia}  \quad {\bf Balaji Krishnamurthy} \quad {\bf Nikaash Puri}\\
{ Media and Data Science Research Lab, Adobe, India} \\
\texttt{\{sshahid, tana, srikanth, sumit.bhatia, kbalaji, nikpuri\}@adobe.com}}

\begin{document}

\maketitle

\begin{abstract}
Hierarchical Topic Models (HTMs) are useful for discovering topic hierarchies in a collection of documents. However, traditional HTMs often produce hierarchies where lower-level topics are unrelated and not specific enough to their higher-level topics. Additionally, these methods can be computationally expensive. We present \textbf{HyHTM} - a \textbf{Hy}perbolic geometry based \textbf{H}ierarchical \textbf{T}opic \textbf{M}odels - that addresses these limitations by incorporating hierarchical information from hyperbolic geometry to explicitly model hierarchies in topic models. Experimental results with four baselines show that HyHTM can better attend to parent-child relationships among topics. HyHTM produces coherent topic hierarchies that specialise in granularity from generic higher-level topics to specific lower-level topics. Further, our model is significantly faster and leaves a much smaller memory footprint than our best-performing baseline.We have made the source code for our algorithm publicly accessible. \footnote{Our code is released at: \url{https://github.com/simra-shahid/hyhtm}}
\end{abstract}

\section{Introduction} \label{sect:introduction}

The topic model family of techniques is designed to solve the problem of discovering human-understandable topics from unstructured corpora \cite{paul2014discovering} where a topic can be interpreted as a probability distribution over words \cite{blei2003latent}. Hierarchical Topic Models (HTMs), in addition, organize the discovered topics in a hierarchy, allowing them to be  compared with each other. The topics at higher levels are generic and broad while the topics lower down in the hierarchy are more specific \cite{teh2004sharing}.  

\begin{figure}[h!]
\centering
\includegraphics[width=0.5\textwidth]{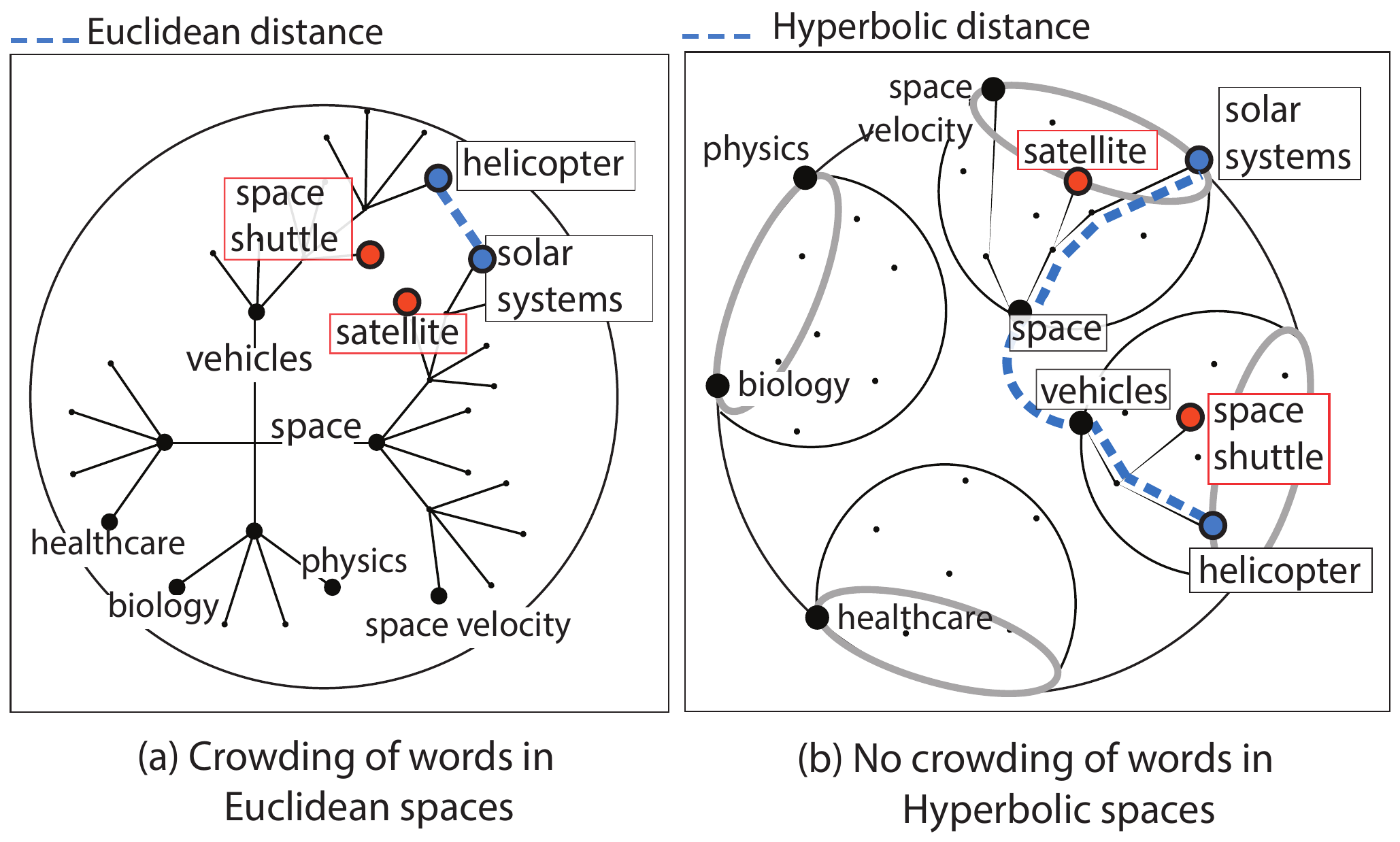} 
\vspace{-15pt}
\caption{In figure (a) we see a concept tree in Euclidean spaces. Words such as \textit{space shuttle} and \textit{satellite}, which belong to moderately different super-concepts such as \textit{vehicles} and \textit{space}, respectively, are brought closer together due to their semantic similarity. This leads to a convergence of their surrounding words, such as \textit{helicopter} and \textit{solar system}, creating a false distance relationship and a crowding effect in Euclidean spaces. In figure (b), we see a concept tree in Hyperbolic spaces (Poincaré ball), which inherently has more space (represented by grey circles) than Euclidean spaces. The distances here grow exponentially towards the edge of the ball, and the concepts at deeper levels such as \textit{helicopter} and \textit{solar systems} move apart in these growing spaces and are far from each other. The dashed blue line shows how the distances in both spaces are calculated.}
\vspace{-10pt}
\label{fig:crowding}
\end{figure}

While significant efforts have been made to develop HTMs \cite{blei2003hierarchical, chirkova2016additive, isonuma2020tree, viegas2020cluhtm}, there are still certain areas of improvement. First, the ordering of topics generated by these approaches provides little to no information about the granularity of concepts within the corpus. By granularity, we mean that topics near the root should be more generic, while topics near the leaves should be more specific.  Second, the lower-level topics must be related to the corresponding higher-level topics. Finally, some of these approaches such as CluHTM \cite{viegas2020cluhtm} are very computationally intensive. We argue that these HTMs have such shortcomings primarily because they do not explicitly account for the hierarchy of words between topics. 

Most of the existing approaches use document representations that employ word embeddings from euclidean spaces. These spaces tend to suffer from the \textbf{crowding problem} which is the tendency to accommodate moderately distant words close to each other \cite{van2008visualizing}. There are several notable efforts that have shown that Euclidean spaces are suboptimal for embedding concepts in hierarchies such as trees, words, or graph entities \cite{chami2019hyperbolic, chami2020low, guo2022co}. In figure \ref{fig:crowding}(a), we show the crowding of concepts in euclidean spaces. Words such as space shuttle and satellite, which belong to moderately different concepts such as vehicles and space, respectively, are brought closer together due to their semantic similarity. This also leads to a convergence of their surrounding words, such as helicopter and solar system creating a false distance relationship. As a result of this crowding, topic models such as CluHTM that use Euclidean word similarities in their formulation tend to mix words that belong to different topics.

Contrary to this, hyperbolic spaces are naturally equipped to embed hierarchies with arbitrarily low distortion \cite{nickel2017poincare, tifrea2018poincar, chami2020low}. The way distances are computed in these spaces are similar to tree distances, i.e., children and their parents are close to each other, but leaf nodes in completely different branches of the tree are very far apart \cite{chami2019hyperbolic}. In figure \ref{fig:crowding}(b), we visualise this intuition on a Poincaré ball representation of hyperbolic geometry (discussed in detail in Section \ref{sect:preliminaries}). As a result of this tree-like distance computation, hyperbolic spaces do not suffer from the crowding effect and words like helicopter and satellite are far apart in the embedding space.

Inspired by the above intuition and to tackle the shortcomings of traditional HTMs, we present \textbf{HyHTM}, a \textbf{Hy}perbolic geometry based \textbf{H}ierarchical \textbf{T}opic \textbf{M}odel which uses hyperbolic geometry to create topic hierarchies that better capture hierarchical relationships in real-world concepts. To achieve this, we propose a novel method of incorporating semantic hierarchy among words from hyperbolic spaces and encoding it explicitly into topic models. This encourages the topic model to attend to parent-child relationships between topics. 

Experimental results and qualitative examples show that incorporating hierarchical information guides the lower-level topics and produces coherent, specialised, and diverse topic hierarchies (Section \ref{sect:experiments}). Further, we conduct ablation studies with different variants of our model to highlight the importance of using hyperbolic embeddings for representing documents and guiding topic hierarchies (Section \ref{sect:roleofHyperbolic}). We also compare the scalability of our model with different sizes of datasets and find that our model is significantly faster and leaves much smaller memory footprint than our best-performing baseline (Section \ref{sect:speed}). We also present qualitative results in Section \ref{sect:quality}), where we observe that HyHTM topic hierarchies are much more related, diverse and specialised. Finally, we discuss and perform in-depth ablations to show the role of hyperbolic spaces and importance of every choice we made in our algorithm (See Section \ref{sect:roleofHyperbolic}).

\begin{figure}
\includegraphics[width=0.5\textwidth]{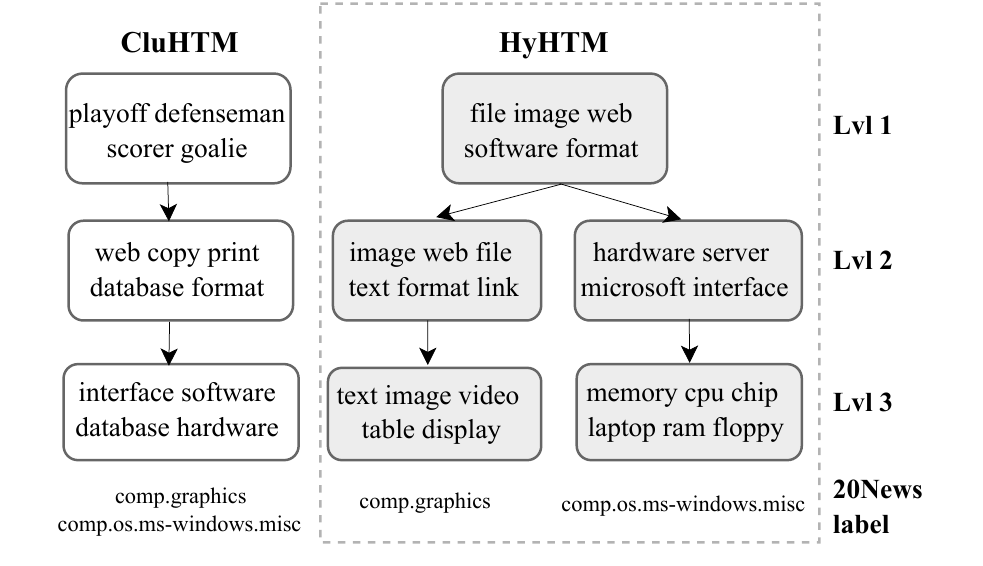} 
\caption{Comparing our Hyperbolic-based HyHTM model to the Euclidean-based CluHTM, for selected 20News document labels (comp.graphics, comp.os.ms-windows.misc), we find that HyHTM is better at discriminating between similar document labels. CluHTM's root-level topics are not related to computer concepts, and it cannot separate these labels at lower levels. HyHTM groups them in the same root level and separates them into different lower-level topics, showing the advantage of using hyperbolic embeddings over euclidean ones to avoid the crowding problem. We show the top words with the highest probability for the topics.}
\label{fig:topic_documentLABEL}
\end{figure}

\section{Related Work} 

To the best of our knowledge, HTMs can be classified into three categories, \textbf{(I) Bayesian generative model} like hLDA \cite{blei2003hierarchical}, and its variants \cite{paisley2013nested, kim2012modeling, tekumalla2015nested} utilize bayesian methods like Gibbs sampler for inferring latent topic hierarchy. These are not scalable due to the high computational requirements of posterior inference. \textbf{(II) Neural topic models} like TSNTM \cite{isonuma2020tree} and others \cite{ wang2021layer, pham2021neural} use neural variational inference for faster parameter inference and some heuristics to learn topic hierarchies but lack the ability to learn appropriate semantic embeddings for topics. Along with these methods, there are \textbf{(III) Non-negative matrix factorization (NMF)} based topic models, which decompose a term-document matrix (like bag-of-words) into low-rank factor matrices to find latent topics. The hierarchy is learned using some heuristics \cite{HSOC, liu2018topic} or regularisation methods \cite{chirkova2016additive} based on topics in the previous level. 

However, the sparsity of the BoW representation for all these categories leads to incoherent topics, especially for short texts. To overcome this, some approaches have resorted to incorporating external knowledge from knowledge bases (KBs) \cite{duan2021topicnet, wang2022knowledge} or leveraging word embeddings \cite{meng2020hierarchical}. Pre-trained word embeddings are trained on a large corpus of text data and capture the relationships between words such as semantic similarities, and concept hierarchies. These are used to guide the topic hierarchy learning process by providing a semantic structure to the topics. \citet{viegas2020cluhtm} utilizes euclidean embeddings for learning the topic hierarchy. However, \citet{tifrea2018poincar, nickel2017poincare, chami2020low, dai2020apo} have shown how the crowding problem in Euclidean spaces makes such spaces suboptimal for representing word hierarchies. These works show how Hyperbolic spaces can model more complex relationships better while preserving structural properties like concept hierarchy between words.  Recently, \citet{shihyperminer} made an attempt to learn topics in hyperbolic embedding spaces. Contrary to the HTMs above, this approach adopts a bottom-up training where it learns topics at each layer individually starting from the bottom, and then during training leverages a topic-linking approach from \citet{duan2021sawtooth}, to link topics across levels. They also have a supervised variant that incorporates concept hierarchy from KBs.

Our approach uses latent word hierarchies from pretrained hyperbolic embeddings to learn the hierarchy of topics that are related, diverse, specialized, and coherent.

\section{Preliminaries}
\label{sect:preliminaries}

We will first review the basics of Hyperbolic Geometry and define the terms used in the remainder of this section. We will then describe the basic building blocks for our proposed solution, followed by a detailed description of the underlying algorithm.

\subsection{Hyperbolic Geometry}
\label{sect:hypgeom}

Hyperbolic geometry is a non-Euclidean geometry with a constant negative Gaussian curvature. Hyperbolic geometry does not satisfy the parallel postulate of Euclidean geometry. Consequently, given a line and a point not on it, there are at least two lines parallel to it. There are many models of hyperbolic geometry, and we direct the interested reader to an excellent exposition of the topic by \citet{book-hyperbolic}. We base our approach on the \textbf{Poincaré ball} model, where all the points in the geometry are embedded inside an $n$-dimensional unit ball equipped with a metric tensor~\citep{nickel2017poincare}. Unlike Euclidean geometry, where the distance between two points is defined as the length of the line segment connecting the two points, given two points $u \in \mathbb{D}^n$ and $v \in \mathbb{D}^n$, the distance between them in the Poincaré model is defined as follows:

\begin{equation} \label{distpoincare}
    \begin{small}
        \begin{multlined}
            d_{P}(u, v) = \operatorname{arcosh} \left(1+ 2 \frac{{\|u-v\|}^{2}}{ (1-\|u\|^{2}) (1-\|v\|^{2})}\right)
        \end{multlined}
    \end{small}
\end{equation}

Here, $\operatorname{arcosh}$ is the inverse hyperbolic cosine function, and $\|.\|$ is the Euclidean norm. Figure~\ref{fig:crowding} has shown an exemplary visualization of how words get embedded in hyperbolic spaces using the Poincaré ball model. As illustrated in Figure~\ref{fig:crowding}(b), distances in hyperbolic space follow a \textit{tree-like} path, and hence they are informally also referred to as \textbf{tree distances}. As can be observed from the figure, the distances grow exponentially larger as we move toward the boundary of the Poincaré ball. This alleviates the crowding problem typical to Euclidean spaces, making hyperbolic spaces a natural choice for the hierarchical representation of data.

\subsection{Matrix Factorization for Topic Models}
\label{sec:nmf}

A \textit{topic} can be defined as a ranked list of strongly associated terms representative of the documents belonging to that topic. Let us consider a document corpus $\mathcal{D}$ consisting of $n$ documents $d_1, d_2, \dots, d_n$, and let $\mathcal{V}$ be the corpus vocabulary consisting of $m$ distinct words $w_1, w_2, \dots, w_m$. The corpus can also be represented by a document-term matrix $\textbf{A} \in \mathbb{R}^{n \times m}$ such that $\textbf{A}_{ij}$ represents the relative importance of word $w_j$ in document $d_i$ (typically represented by the TF-IDF weights of $w_i$ in $d_j$).

A popular way of inferring topics from a given corpus is to factorize the document-term matrix. Typically, non-negative Matrix Factorization (NMF) is employed to decompose the document-term matrix, $\textbf{A}$, into two non-negative approximate factors: $\textbf{W} \in \mathbb{R}^{n \times \textbf{N}}$ and $\textbf{H} \in \mathbb{R}^{\textbf{N} \times m}$. Here, $\textbf{N}$ can be interpreted as the number of underlying topics. The factor matrix $\textbf{W}$ can then be interpreted as the document-topic matrix, providing the topic memberships for documents, and $\textbf{H}$, the topic-term matrix, describes the probability of a term belonging to a given topic. This basic algorithm can also be applied recursively to obtain a hierarchy of topics by performing NMF on the set of documents belonging to each topic produced at a given level to get more fine-grained topics \citep{chirkova2016additive, viegas2020cluhtm}.
\section{Hierarchical Topic Models Using Hyperbolic Geometry} 
\label{sect:proposed-work}
We now describe HyHTM -- our proposed Hyperbolic geometry-based Hierarchical Topic Model. We first describe how we capture semantic similarity and hierarchical relationships between terms in hyperbolic space. We then describe the step-by-step algorithm for utilizing this information to generate a topic hierarchy.

\subsection{Learning Document Representations in Hyperbolic Space and Root Level Topics}
\label{sect:mining}

As discussed in Section~\ref{sec:nmf}, the first step in inferring topics from a corpus using NMF is to compute the document-term matrix $\textbf{A}$. A typical way to compute the document-term matrix $\textbf{A}$ is by using the TF-IDF weights of terms in a document that provides reprsentations of the documents in the term space. However, usage of TF-IDF (and its variants) results in sparse representations and ignores the semantic relations between different terms by considering only the terms explicitly present in a given document. \citet{viegas2019cluwords} proposed an alternative formulation for document representations that utilizes pre-trained word embeddings to enrich the document representations by incorporating weights for words that are semantically similar to the words already present in the document. The resulting document representations are computed as follows.
\begin{equation} \label{tfidf}
    \textbf{A} = ( \textbf{TF}  \times \textbf{M}_\text{S} ) \odot ( \mathbf{1} \times \textbf{IDF}^T)
\end{equation}


Here, $\odot$ indicates the Hadamard product. $\textbf{A}$ is the $n \times m$ document-term matrix. $\textbf{TF}$ is the term-frequency matrix such that $\textbf{TF}_{i,j} = tf(d_i, w_j)$ and $\textbf{M}_\text{S}$ is the $m \times m$ term-term similarity matrix that captures the pairwise semantic relatedness between the terms and is defined as $\textbf{M}_{\text{s}_{i,j}} = sim(w_i, w_j)$, where $sim(w_i, w_j)$ represents the similarity between terms $w_i$ and $w_j$ and can be computed using typical word representations such as word2vec~\citep{w2v} and GloVe~\citep{pennington2014glove}. Finally, \textbf{IDF} is the $m \times 1$ inverse-document-frequency vector representing the corpus-level importance of each term in the vocabulary. Note that \citet{viegas2019cluwords} used the following modified variant of IDF in their formulation, which we also chose in this work.

\begin{equation}
\begin{small}
\textbf{IDF}(i) = \log \left(  \frac{|D|}{\sum_{d \in D} \mu \left( w_i, d \right)}\right)
\end{small}
\end{equation}
Here, $\mu \left( w_i, d \right)$ is the average of the similarities between  term $w_i$ and all the terms $w$ in document $d$  such that $\textbf{M}_\text{S}(w_i, w) \neq 0$. Thus, unlike traditional IDF formulation where the denominator is document-frequency of a term, the denominator in the above formulation captures the semantic contribution of $w_i$ to all the documents.

In our work, we adapt the above formulation to obtain document representations in Hyperbolic spaces by using \textbf{Poincaré GloVe embeddings}~\citep{tifrea2018poincar}, an extension of the traditional Euclidean space GloVe~\citep{pennington2014glove} to hyperbolic spaces. Due to the nature of the Poincaré Ball model, the resulting embeddings in the hyperbolic space arrange the correspondings words in a hierarchy such that the sub-concept words are closer to their parent words than the sub-concept words of other parents.

There is one final missing piece of the puzzle before we can obtain suitable document representations in hyperbolic space. Recall that due to the nature of the Poincaré Ball model, despite all the points being embedded in a unit ball, the hyperbolic distances between points, i.e., tree distances (Section~\ref{sect:hypgeom}) grow exponentially as we move towards boundary of the ball (see Figure~\ref{fig:crowding}). Consequently, the distances are bounded between $0$ and $1$. As NMF requires all terms in the input matrix to be positive, we cannot directly use these distances to compute the term-term similarity matrix $\textbf{M}_\text{S}$ in Equation~\eqref{tfidf} as $1-d_{P}\left(w, w'\right)$ can be negative. To overcome this limitation, we introduce the notion of \textbf{Poincaré Neighborhood Similarity}, ($s_{p_n}$), which uses a neighborhood normalization technique. The $k$-neighborhood of a term $w$ is defined as the set of top k-nearest terms ${w_{1}, ..., w_{k}}$ in the hyperbolic space and is denoted as $n_k(w)$. For every term in the vocabulary $\mathcal{V}$, we first calculate the pair-wise poincaré distances with other terms using Equation~\eqref{distpoincare}. Then, for every term $w \in \mathcal{V}$, we compute similarity scores with all the other terms in its $k$-neighborhood $n_k(w)$ by dividing each pair-wise poincaré distance between the term and its neighbor by the maximum pair-wise distance in the neighborhood. This can be represented by the following equation where $w' \in n_{k}(w)$:
\useshortskip
\begin{equation}
\label{eq:spn}
    s_{p_n}\left(w, w'\right)=1-\frac{d_{P}\left(w, w'\right)}{\max\limits_{w_a, w_b \in n_{k}(w) }(d_{P}\left(w_a, w_b\right) )}
\end{equation}

With this, we can now compute the term-term similarity matrix $\boldsymbol{M_{S}}$ as follows.
\useshortskip
\begin{equation}\label{similarity} 
\small
    \textbf{M}_\text{S}{(w, w')}=\begin{cases} s_{p_n}\left(w, w'\right) & \text { if } s_{p_n}\left(w, w'\right) \geq \alpha, \\
    0 & \text { otherwise } \end{cases}
\end{equation}
Note that there are two hyperparameters to control the neighborhood -- \textit{(i)} the neighborhood size using $k_{s}$; and \textit{(ii)} the quality of words using $\alpha$, which keeps weights only for the pair of terms where the similarity crosses the pre-defined threshold $\alpha$ thereby reducing noise in the matrix. Without $\alpha$, words with very low similarity may get included in the neighborhood eventually leading to noisy topics. 

We now have all the ingredients to compute the document-representation matrix $\textbf{A}$ in the hyperbolic space and NMF can be performed to obtain the first set of topics from the corpus as described in Section~\ref{sec:nmf}. This gives us the \textit{root} level topics of our hierarchy. Next, we describe how we can discover topics at subsequent levels.

\subsection{Building the Topic Hierarchy}
In order to build the topic hierarchy, we can iteratively apply NMF for topics discovered at each level as is typically done in most of the NMF based approaches. However, do note that working in the Hyperbolic space allows us to utilize hierarchical information encoded in the space  to better guide the discovery of topic hierarchies. Observe that the notion of similarity in the hyperbolic space as defined in Equation\eqref{eq:spn} relies on the size of the neighborhood. In large neighborhood, a particular term will include not only its immediate children and ancestors but also other semantically similar words that may not be hierarchically related. On the other hand, a small neighborhood will include only the immediate parent-child relationships between the words, since subconcept words are close to their concept words. HyHTM uses this arrangement of words in hyperbolic space to explicitly guide the lower-level topics to be more related and specific to higher-level topics. In order to achieve this, we construct a \textbf{Term-Term Hierarchy} matrix,  $\textbf{M}_\text{H} \in R^{|V|\times|V|}$ as follows.
\useshortskip
\begin{equation} 
\label{hierarchy}
\small
    \textbf{M}_\text{H}{(w, w')}=\begin{cases}
    1 & \text {if }  w' \in n_{k_{h}}(w),\\
    0 & \text { otherwise }
    \end{cases}
\end{equation}
Here, $k_{H}$ is a hyperparameter that controls the neighborhood size. $\textbf{M}_\text{H}$ is a crucial component of our algorithm as it encodes the hierarchy information and helps guide the lower-level topics to be related and specific to the higher-level topics.

Without loss of generality, let us assume we are at $i^{th}$ topic node $t_{i}$ at level $l$ in the hierarchy. We begin by computing $\textbf{A}_0 = \textbf{A}$, as outlined in Equation~\eqref{tfidf}, at the root node (representing all topics) and subsequently obtaining the first set of topics (at level $l=1$). Also, let the number of topics at each node in the hierarchy be $N$ (a user-specified parameter). Every document is then assigned to one topic with which it has the highest association in the document-topic matrix $\textbf{W}_{l-1}$. Once all the documents are collected into disjoint parent topics, we use a subset of $\textbf{A}_0$ with only the set of documents ($\mathcal{D}_{t_j}$) belonging to the $j^{th}$ topic, and denote this by $\textbf{A}_{l-1}$. We then branch out to $N$ lower-level topics at the $i^{th}$ node, using the following steps:

\textbf{Parent-Child Re-weighting for Topics in the Next Level}: We use the term-term hierarchical matrix $\textbf{M}_\text{H}$ to assign more attention to words hierarchically related to all the terms in the topic node $t_{i}$, and guide the topic hierarchy so that the lower-level topics are consistent with their parent topics. We take the product of the topic-term matrix of the $t_{i}$, denoted by, $\textbf{H}_{i}$  with the hierarchy matrix $\textbf{M}_\text{H}$. This assigns weights with respect to associations in the topic-term matrix
\begin{equation}\label{eq7}
\textbf{M}_{ti} = \mathbbm{1}_i^T\textbf{H}_{l-1} \times \textbf{M}_\text{H}
\end{equation}
Here, $\mathbbm{1}_i$ is the one-hot vector for topic $i$, and $\textbf{H}_{l-1}$ is the topic-term factor obtained by factorizing the document-representations $\textbf{A}_{l-1}$ of the parent level.

\textbf{Document representation for computing next level topics}: We now compute the updated document representations for documents in topic node $t_{i}$ that infuse semantic similarity between terms with hierarchical information as follows.
\useshortskip
\begin{equation} \label{combine-final}
\textbf{A}_l = \textbf{A}_{l-1} \odot  \textbf{M}_{ti} 
\end{equation}

By using the updated document representations $\textbf{A}_l$  we perform NMF as usual and obtain topics for level $l+1$. The algorithm then continues to discover topics at subsequent levels and stops exploring topic hierarchy under two conditions -- \textit{(i)} if it reaches a topic node such that the number of documents in the node is less than a threshold ($D_{min}$); \textit{(ii)} when the maximum hierarchy depth ($\mathcal{L}_{max}$) is reached. We summarize the whole process in the form of a pseudcode in Algorithm \ref{alg:one}. 

\RestyleAlgo{ruled}

\SetKwComment{Comment}{/* }{ */}
\IncMargin{1em}
\begin{algorithm}
\caption{\textbf{The HyHTM Algorithm}}\label{alg:one}
\SetKwFunction{gethtopics}{\textsc{GetHier}}
\SetKwFunction{nmf}{\textsc{NMF}}
\SetKwFunction{ghm}{\textsc{Get-Hier-Matrix}}
\SetKwFunction{dot}{\textsc{Dot}}
\SetKwFunction{gt}{\textsc{Get-Topic-Words}}
\SetKwFunction{gd}{\textsc{Get-Topic-Docs}}
\SetKwFunction{ah}{\textsc{Add-To-Hierarchy}}

\SetKwInOut{Input}{Input}
\SetKwInOut{Output}{Output}
\Input{
Max depth level ($\mathcal{L}_{\text{max}}$) \\
Min \# of documents ($D_{\text{min}}$)  \\
Default \# of topics ($N$) }
\Output{Hierarchy of Topics}
Compute $\textbf{A}$ using Eq \eqref{tfidf} \& \eqref{similarity}\\
\gethtopics{$\textbf{A}$,  $1$} \\
\textbf{def} \gethtopics{$\textbf{A}$,  $L$}:

\Indp
\textbf{if} { $L$ $>$  $L_{max}$ \text{or len}($\textbf{A}$) $<$ $D_{min}$}: return\\
$\textbf{W}_{l-1}, \textbf{H}_{l-1} \gets \nmf(\textbf{A}, N) $ \\
\For{$i = 0$ to $H_{l-1}.\text{size}$}{

    Get parent topic using $\textbf{H}_{l-1}$\\
    Add topic to hierarchy \\

    Get Docs of topic $t_j$ using $\textbf{W}_{l-1}$\\
    Get $\textbf{A}_{l-1}$ for $D_{t_j}$ from $\textbf{A}_0$ \\
    
    Compute Parent-Child Reweighting $\textbf{M}_{ti}$ using Eq \eqref{eq7}\\

    Compute $\textbf{A}_{l}$ next level from $\textbf{M}_{ti}$ \& $\textbf{A}_{l-1}$ using Eq \eqref{combine-final}\\
 
  \gethtopics{$\textbf{A}_{l}$, $L+1$}
  }
\end{algorithm}

\section{Experimental Setup} 

\noindent\textbf{Datasets:}\label{datasets} To evaluate our topic model, we consider 8 well-established public benchmark datasets. In Table \ref{tab:dataset} we report the number of words and documents, as well as the average number of words per document. We have used datasets with varying numbers of documents and average document lengths. We provide preprocessing details in the Appendix (See \ref{sect:preprocessing}).

\begin{table}[h!]
\begin{adjustbox}{max width=0.5\textwidth} 
\begin{tabular}{lccc} 
 \toprule
 Dataset & Vocabulary & No. of Documents & \multicolumn{1}{p{1.5cm}}{ \centering Avg. Doc \\ Length }
 \\ 
 \midrule
 
InfoVis-Vast (\textbf{\href{https://www.cc.gatech.edu/gvu/ii/jigsaw/datafiles.html}{InfoVAST}}) & 8,309 & 1,085 & 153.62 \\
\href{https://www.cc.gatech.edu/gvu/ii/jigsaw/datafiles.html}{Neurips}  & 9,407  & 1,499 & 517.9 \\

\href{http://mlg.ucd.ie/datasets/bbc.html}{BBC} & 6,384 & 2,255 & 209.00 \\

 20Newsgroup (\textbf{\href{http://qwone.com/~jason/20Newsgroups/}{20News}}) & 12,199  & 18,846 & 119.80 \\
Enron     & 10,116  & 39,860 & 93.29 \\
Amazon Reviews \textbf{(\href{https://www.kaggle.com/kashnitsky/hierarchical-textclassification/version/1}{Amazon})}   & 9,458 & 40,000 & 39.04 \\
Web of Science \textbf{(WOS)}   & 40,755 & 46,985 & 132.30 \\
 \href{http://groups.di.unipi.it/gulli/AG_corpus_of_news_articles.html}{AGNews}  & 17,436 & 127,600 & 24.15 \\
 \hline
\end{tabular}
\end{adjustbox}
\caption{Dataset characteristics}
\label{tab:dataset}
\end{table}

\noindent\textbf{Baseline Methods:} Our model is a parametric topic model which requires a fixed number of topics to be specified. This is different from non-parametric models, which automatically learn the number of topics during training. For the sake of completeness, we also compare our model to various non-parametric models such as \textbf{hLDA} \cite{blei2003hierarchical} a bayesian generative model, and \textbf{TSNTM} \cite{isonuma2020tree} which uses neural variational inference. We also compare with NMF-based parametric models like \textbf{hARTM }\cite{chirkova2016additive} which learns a topic hierarchy with a bag of words of documents and \textbf{CluHTM }\cite{viegas2020cluhtm} which uses euclidean based pre-trained embeddings \cite{fasttext} to provide semantic similarity context to topic models. We provide the implementation details of these baselines in the Appendix (See \ref{sect:implementation_details}).

\noindent\textbf{Number of topics:} hARTM only allows fixing the total number of topics at a level and cannot specify the number of child topics for every parent topic. CluHTM, on the other hand, has a method to learn the optimal number of topics, but it is highly inefficient\footnote{The training time of CluHTM 20News was approximately 32 hours, and for Amazon was approximately 22 hours. For every branch and level, it runs an empirical analysis for topics in ranges 5 and 20 and picks the topic number corresponding to the best coherence.}. We use the same number of topics for fair comparison in hARTM, CluHTM, and HyHTM. We fix the number of topics for the top level as 10, with 10 sub-topics under each parent topic. The total number of topics at each level is 10, 100, and 1000.  Non-parametric models hLDA and TSNTM learn the number of topics, and we report these numbers in the appendix (See \ref{sect:numberOfTopics_nonparametric}).

\noindent We select the best values for the hyperparameters $k_H$, $k_S$, and $\alpha$ by tuning them for the model with the best empirical results. We report these in the Appendix \ref{sect:implementation_details}.

\section{Experimental Results}\label{sect:experiments}

In this section we compare our model's performance on well-estabilished metrics to assess the coherence, specialisation, and diversity of topics. We present qualitative comparision for selected topics in Figure \ref{fig:topic_documentLABEL} and in Appendix \ref{sect:quality}. We discuss and perform ablations to show the role of hyperbolic spaces and effectiveness of our algorithm (See Appendix \ref{sect:roleofHyperbolic}).

\noindent \textbf{RQ1: Does HyHTM produce coherent topics?} 
Topic coherence is a measure that can be used to determine how much the words within a topic co-occur in the corpus. The more the terms co-occur, the easier it is to understand the topic. We employ the widely used coherence measure from \citet{aletras2013evaluating} and report the average across the top 5 and 10 words for every topic in Table \ref{table:1}. We observe that for majority of the datasets, HyHTM consistently ranks at the top or second highest in terms of coherence. We also observe that for some cases hLDA and TSNTM, which have very few topics (See \ref{sect:numberOfTopics_nonparametric}) compared to HyHTM, have higher coherence values. To this end, we conclude that incorporating neighborhood properties of words from hyperbolic spaces can help topic models to produce topics that are comprehensible and coherent. Coherence is mathematically defined as,

\begin{equation} \label{coherence}
\text{Coherence} = \frac{{\sum_{{i=1}}^{n-1}{\sum_{{j=i+1}}^{n} \log \frac{{P(w_i, w_j)}}{{P(w_i) P(w_j)}}}}}{{\binom{n}{2}}}
\end{equation}

where $w_i$ and $w_j$ are words in the topic, while $P(w_i, w_j)$ and $P(w_j)$ are the probabilities of co-occurrence of $w_i$ and $w_j$ and the of occurrence of $w_j$ in the corpus respectively.

\begin{table}[h!]
\centering
\begin{adjustbox}{max width=0.5\textwidth} 
\begin{tabular}{lcccc|c}
\toprule
Dataset &  hLDA & TSNTM & hARTM & CluHTM & \textbf{HyHTM} \\
\midrule
InfoVAST & \textbf{0.061} & 0.017 & 0.044 & 0.027 & \underline{0.045} \\
Neurips  & 0.066 & 0.133 & 0.084  & \underline{0.226} & \textbf{0.338}   \\
{BBC}  & 0.232 & \underline{0.248} & \textbf{0.296}  & 0.181  & 0.235 \\
{20News}   & 0.214 & 0.279 & \textbf{0.325}  & 0.293 & \textbf{0.325} \\
{Enron} & 0.226 & 0.250 & 0.327 & \underline{0.346} & \textbf{0.365}  \\
{Amazon}  & 0.127 & 0.097 & \textbf{0.166} & 0.124 & \underline{0.158}   \\
{WOS}  & 0.024 & \textbf{0.096} & 0.025 & 0.010 & \underline{0.052} \\
{AGNews} & 0.145 & \textbf{0.209} & 0.142 & 0.039 & \underline{0.154} \\

\bottomrule
\end{tabular}
\end{adjustbox}
\caption{Comparing topic coherence, where higher coherence is better. Bold represents the best-performing metric and underline represents the second-best metric.} 
\label{table:1}
\end{table}

\noindent \textbf{RQ2: Does HyHTM produce related and diverse hierarchies?} To assess the relationships between higher-level parent topics and lower-level child topics, we use two metrics: (i) hierarchical coherence, and (ii) hierarchical affinity. 

\noindent \textbf{Hierarchical Coherence}: We build upon the coherence metric above to compute the coherence between parent topic words and child topic words. For every parent-topic and child-topic pair, we calculate the average across the top 5 words and top 10 words and report this in Table \ref{table:2}. We observe that HyHTM outperforms the baselines across datasets, and we attribute this result to our parent-child reweighting framework of incorporating the hierarchy of higher-level topics. In most cases, hLDA and TSNTM have very low hierarchical coherence because the topics generated by these models are often too generic across levels and contain multiple words from different concepts, whereas hARTM and CluHTM have reasonable scores and are often better than these.  From this observation, we conclude that adding hierarchies from hyperbolic spaces to topic models produces a hierarchy where lower-level topics are related to higher-level topics. Hierarchical coherence is defined as,
\begin{equation} \label{hcoherence}
\text{HCoherence} = \frac{{\sum_{{i=1}}^{n}{\sum_{{j=1}}^{n} \log \frac{{P(w_i, w_j)}}{{P(w_i) P(w_j)}}}}}{{n^2}}
\end{equation}
where $w_i$ and $w_j$ represent words from the parent topic and child topic, while $P(w_i, w_j)$ and $P(w_j)$ are the probabilities of co-occurrence of $w_i$ and $w_j$ and the of occurrence of $w_j$ in the corpus respectively.

\begin{table}[h!]
\centering
\begin{adjustbox}{max width=0.5\textwidth} 
\begin{tabular}{lcccc|c}
\toprule
\cmidrule(l){2-6} 
Dataset &  hLDA & TSNTM & hARTM & CluHTM & \textbf{HyHTM} \\
\Xhline{3\arrayrulewidth}
{InfoVAST}   & 0.011  & \underline{0.018} & 0.007  & 0.011 & \textbf{0.025} \\
{Neurips}   & 0.059 &	0.019	 & 0.049 & \underline{0.063} & \textbf{0.296} \\
{BBC}   & 0.064  & 0.089 & \underline{0.211} & 0.102 & \textbf{0.221}  \\
{20News}   & 0.031 & 0.049 & \underline{0.133} & 0.127 & \textbf{0.287} \\
{Enron}  & 0.023 & 0.068 & \underline{0.139}  & 0.107 & \textbf{0.329}  \\
{Amazon} & 0.008  & 0.056 & 0.073 & \underline{0.085} & \textbf{0.123}  \\
{WOS}   &  0.006  & \underline{0.022}  &  0.016 & 0.002 & \textbf{0.045}  \\
{AGNews} & 0.017 & 0.018 & 0.046 & \underline{0.071} & \textbf{0.151} \\
\bottomrule
\end{tabular}
\end{adjustbox}
\caption{Comparing Hierarchical Coherence. Bold represents the best-performing metric and underline represents the second-best metric.} 
\label{table:2}
\end{table}

\begin{figure}[h!]
\centering
\includegraphics[width=0.5\textwidth, scale=1]{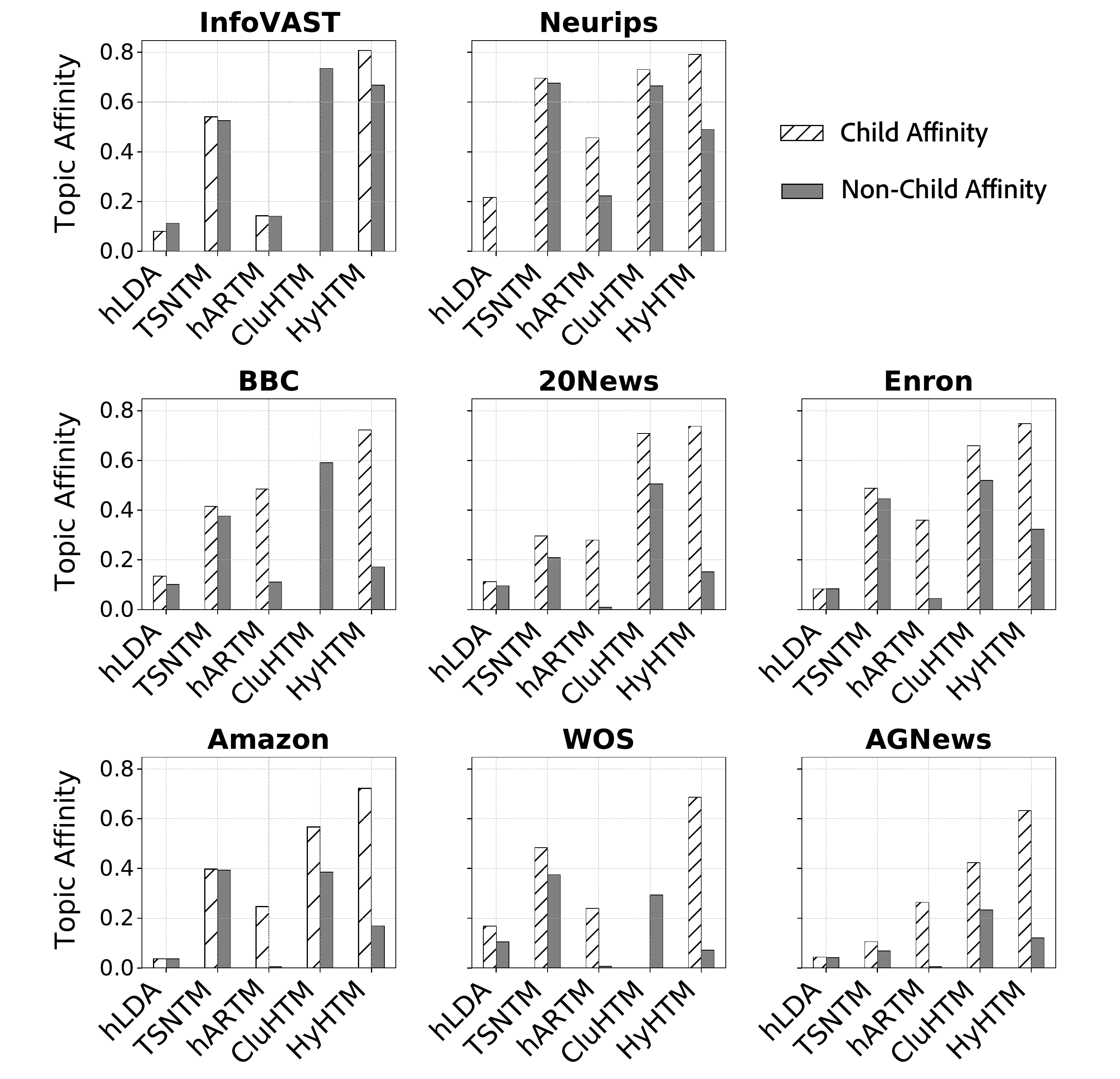}
\caption{Analysis of Hierarchical Topic Affinities. A higher Child Affinity value indicates stronger relatedness between parent and child topics. The more the difference between Child to Non-Child Affinities, the more diverse the topics are in the hierarchy. Please note, some affinities appear to be missing in the visualization due to their significantly lower magnitudes compared to the highest affinity value."}
\label{fig:Hierarchical Affinity subset}
\end{figure}

\noindent \textbf{Hierarchical Affinity}: We employ this metric from \citet{isonuma2020tree} which considers the topics at levels 2 as parent topics and the topics at level 3 to compute (i) \textbf{child affinity}, and, (ii) \textbf{non-child affinity}. The respective affinities are measured by the average cosine similarity of topic-term distributions between parent \& child and parent \& non-child topics. \footnote{Hierarchical Affinity metric is independent of the embedding space the models were they are trained on.} When child affinity is higher than non-child affinity, it implies (i) the topic hierarchy has a good diversity of topics, and, (ii) the parents are related to their children. We present the hierarchical affinities in figure \ref{fig:Hierarchical Affinity subset}.

\noindent We observe that HyHTM has the largest between child affinities across all the datasets. We also observe that the difference between child and non-child affinities is also larger than that for any other baseline. hLDA and TSNTM have very similar child and non-child affinities, which indicates how generic topics are across the hierarchy. In hARTM, we observe high child affinity and negligible non-child affinity. From these observations, we conclude that HyHTM produces related and diverse topics.

\textbf{RQ3: Does HyHTM produce topics with varying granularity across levels?} We use the Topic specialisation metric from \citet{kim2012modeling}, to understand the granularity of topics in the hierarchy. Topic specialization is the cosine distance between the term distribution of a topic with the term distribution of the whole corpus. According to the metric, the root-level topics are trained on the whole corpus so they are very generic, while the lower-level topics are trained on a subset of documents, and they specialise. A higher specialization value means that the topic vector is not similar to the whole corpus vector, and hence it is more specialised. With increasing depth in the hierarchy, the specialisation of a topic should increase and its distance from the corpus vector must increase to model reasonable topic hierarchies described above. 

\begin{figure}[h!]
\includegraphics[width=0.5\textwidth, scale=1]{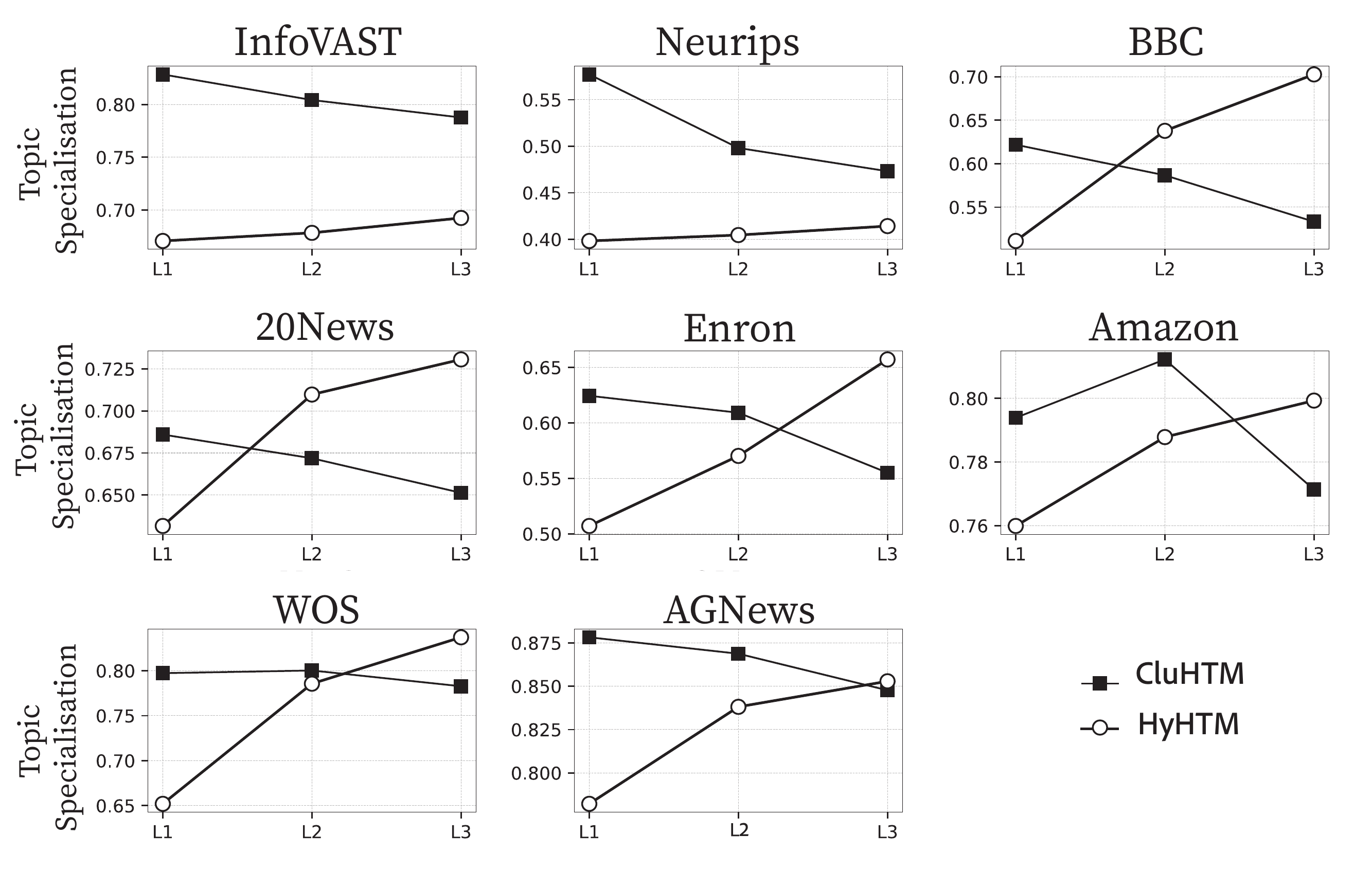}
\caption{Comparision between Topic Specialisation of CluHTM and HyHTM for different datasets. An increasing trend from Level 1 (L1) to Level 3 (L3) indicates that topics are becoming more specific, diverging from a more generic corpus-word distribution.}
\label{fig:complete_topicSpecialisation_all}
\end{figure}

As the resulting topic-proportions and range of topic-specialisation of CluHTM and HyHTM are similar, we first focus on these models to effectively underscore the advantages of employing hyperbolic spaces. As depicted in Figure \ref{fig:complete_topicSpecialisation_all}, unlike CluHTM, our HyHTM model consistently exhibits an increasing trend in topic specialization across majority of the datasets. We attribute this result to our use of hyperbolic spaces in our algorithm which groups together documents of similar concepts from the root level itself.

Additionally, we present the topic specialization of other models in Appendix Table \ref{tab:topic_special_others}. We discover that TSNTM usually scores low, suggesting generic topics at all levels. Although hLDA shows increasing specialization, it seemingly fails to generate related topic hierarchies, as evidenced by quantitative metrics and qualitative topics (See Section \ref{sect:quality}). Despite hARTM showing an increase in granularity, it often lumps unrelated concepts under a single topic hierarchy, akin to CluHTM, as illustrated in the qualitative examples (See Section \ref{sect:quality}).

\subsection{Runtime \& Memory footprint}\label{sect:speed}

\begin{figure}[h!]
\includegraphics[width=0.5\textwidth]{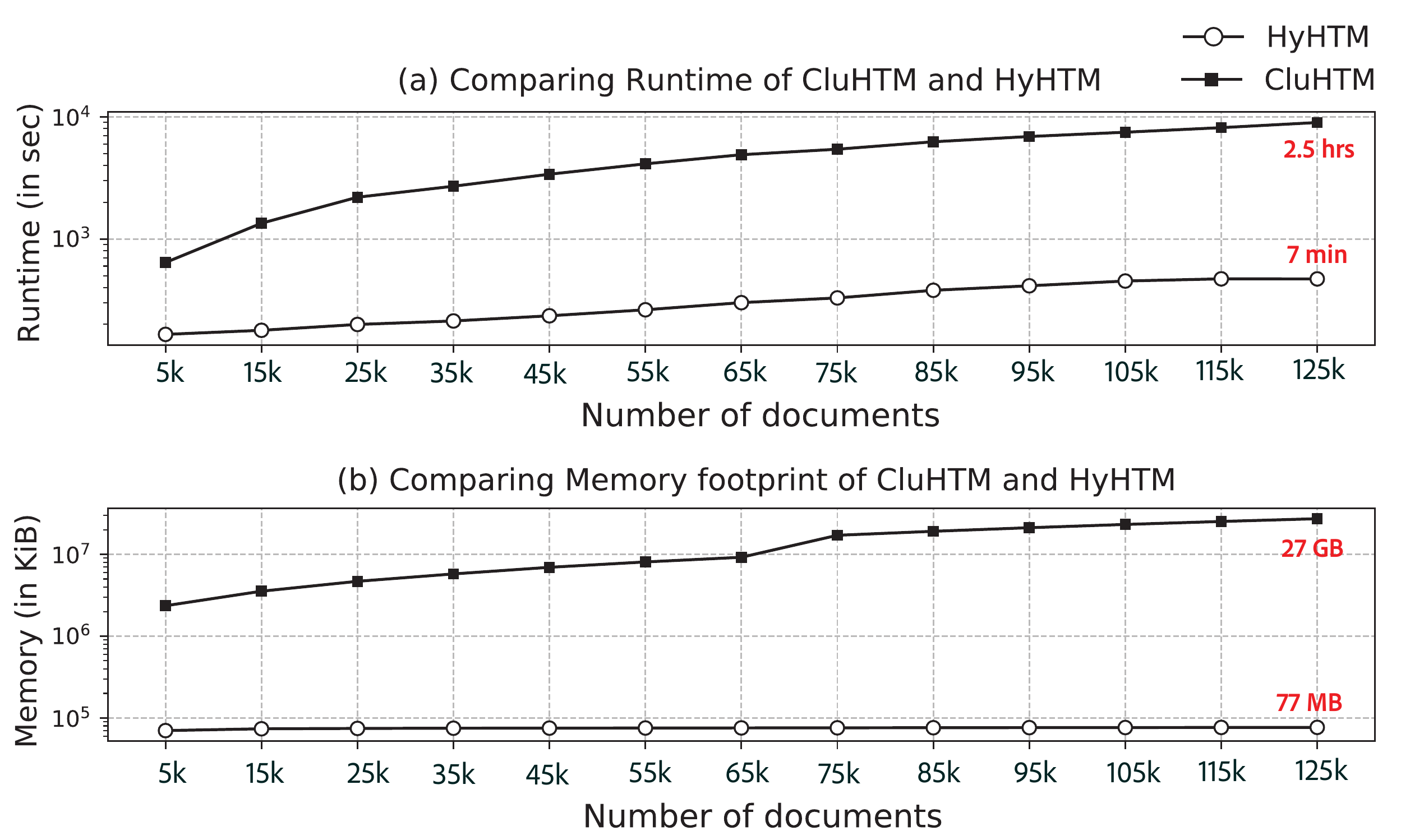}
\caption{Comparing runtime and memory footprint for HyHTM (our model) and CluHTM on AGNews dataset.}
\label{fig:time-memory}
\end{figure}

To evaluate how our model scales with the size of the datasets, we measure the training time and memory footprint by randomly sampling a different number of documents (5k to 125k) from the AGNews dataset. From Figure \ref{fig:time-memory} we observe that, as the number of documents increases, the training time of our model does not change considerably, whereas that of the CluHTM increases significantly. HyHTM can be trained approximately 15 times faster than the CluHTM model with even 125k documents. CluHTM works inefficiently by keeping the document representations of all the topics at a level in the working memory. This is a result of CluHTM developing the topic hierarchy in a breadth-first manner.  We have optimized the HyHTM code to train one branch from root to leaf in a depth-first manner which makes our model more memory and efficient. hLDA took approximately 1.32 hours for training on the complete dataset, and hARTM and TSNTM took more than 6 hours.\\

\subsection{Quality of Topics}\label{sect:quality} 
To intuitively demonstrate the ability of our model to generate better hierarchies, we present topic hierarchies of all models for some selected 20News target labels in the Appendix in Figure \ref{fig:space_hierarchy}. \footnote{We present only those topic-hierarchies where most of the documents of the respective 20News label lies.}  
Across various topic categories, unlike HyHTM, other models tend to struggle with delineating specific subconcepts, maintaining relatedness, and ensuring specialization within their topics, which highlights HyHTM’s improved comprehensibility. 
For the \textit{sci.space} 20News label, we observe that topics from CluHTM across all the levels are related space concepts but it is challenging to label them as specific subconcepts. The hARTM topics for space has a resonable hierarchy but it has documents of different concepts such as \textit{sci.space, sci.med, rec.sports.baseball}.  For hLDA and TSNTM, the lack of relatedness and specialization makes it difficult to identify these topics as space-themed. A similar trend can be observed for \textit{comp.os.ms-windows.misc} and \textit{sci.med} 20News categories in the figure, where the models exhibit similar struggles.

\section{Ablation}\label{sect:roleofHyperbolic}

\noindent \textbf{Do Hyperbolic embeddings represent documents better than Euclidean ones?} 

\noindent To investigate this we consider a variant of our model called \textbf{Ours (Euc)} which incorporates pretrained Fasttext \cite{bojanowski2017enriching} (trained on euclidean spaces) instead of Poincare embeddings in $M_s(w, w')$, and we keep all the other steps unchanged. From Table \ref{table:ablation12}, we observe that using hyperbolic embeddings for guiding parent-child in $A_{l}$ is better choice as it produces topics that are more coherent and hierarchies in which lower-level topics are related to higher-level topics. 

\begin{table}[h!]
\centering
\begin{adjustbox}{max width=0.4\textwidth} 
\begin{tabular}{lcccccc}
\toprule
  & \multicolumn{2}{c}{\textbf{20News}}
  & \multicolumn{2}{c}{\textbf{Amazon}} \\
\cmidrule(r){2-3} \cmidrule(l){4-5}
& { \centering Coh} & { \centering Hier Coh} & { \centering Coh} & { \centering Hier Coh}\\
\midrule
{\small Ours} & \textbf{0.325} & \textbf{0.287} &   \textbf{0.158} & \textbf{0.123} \\
{\small Ours (Euc)} & \underline{0.322} & \underline{0.240}  & \underline{0.156} & \underline{0.113}  \\
{\small  CluHTM } & 0.293 & 0.127 & 0.124 &   0.085\\
\bottomrule
\end{tabular}
\end{adjustbox}
\caption{Analysis the role of hyperbolic embeddings}
\label{table:ablation12}
\end{table}

\noindent \textbf{Does enforcing hierarchy between parent-child topics in equation \ref{combine-final} result in better hierarchy?} 

\noindent We examine this by comparing the \textbf{Ours (Euc)} variant and the CluHTM baseline. Both models use identical underlying document representations, yet they differ in how they guide their hierarchies, particularly in the equation \ref{combine-final} of our model. As demonstrated in Table \ref{table:ablation12}, \textbf{Ours (Euc)}, which accounts for word hierarchies between higher-level and lower-level topics, generates topic hierarchies that are nearly twice as effective in terms of topical hierarchical coherence and hierarchical affinity.

In the Appendix (See Section \ref{sect:ablation}), we also examine the importance of our approach by replacing the underlying algorithm with hierarchical clustering methods.

\section{Conclusion}
In this paper, we have proposed HyHTM, which uses hyperbolic spaces to distill word hierarchies of higher-level topics in order to refine lower-level topics. Both quantitative and qualitative experiments have demonstrated the effectiveness of HyHTM in creating hierarchies in which lower-level topics are realted and more specific than higher-level topics. HyHTM is much more efficient compared to our best-performing baseline. A major limitation of HyHTM is that it is parametric and therefore requires empirical analysis to find the optimal number of topics at each level. We plan to investigate this shortcoming in the future.
\clearpage
\section{Limitations}\label{sect:limitations}

In this paper, we propose a method to effectively incorporate the inherent word hierarchy in topic models for hierarchical topic mining. We use poincare embeddings, trained on wikipedia, to compute the hierarchical relatedness between words. Hence, our model relies on how well these embeddings are trained and whether they effectively capture the word hierarchy. Moreover, any bias in the embeddings is translated into our model. The second major limitation of our model is that since these embeddings are trained on wikipedia, they may not perform well on datasets that are very different from wikipedia or on datasets where the relation between two words is very different from their relation in wikipedia. For example, \textit{topic} and \textit{hierarchy} will have a very different relation in scientific journals from what they have in wikipedia. Our model is parametric HTM, and we plan on investigating methods to induce number of topics using hyperbolic spaces. 

\section{Ethics Statement}\label{sect:ethics}

\begin{itemize}
    \item The dataset used to train the poincare embeddings is Wikipedia Corpus, a publicly available dataset standardized for research works.
    \item We have added references for all the papers, open-source code repositories and datasets.  
    \item In terms of dataset usage for topic modeling, we have used only publicly available datasets. We also ensure that any datasets used in our research do not perpetuate any harmful biases. 
    \item We also plan to make our models publicly available, in order to promote transparency and collaboration in the field of natural language processing.
\end{itemize}
\bibliography{custom}
\bibliographystyle{acl_natbib}

\appendix

\section{Additional Results}

\subsection{Topic Specialisation} \label{sect:completeTS}

In section \ref{sect:experiments} we report the Topic Specialisation for CluHTM and HyHTM. In this section we present the topic specialisation results in the table \ref{tab:topic_special_others}.

\begin{table}[h!]
\centering
\begin{adjustbox}{max width=0.5\textwidth} 
\begin{tabular}{lccc}
\toprule
\textbf{Dataset} & \textbf{Lvl 1} & \textbf{Lvl 2} & \textbf{Lvl 3}\\
\midrule
\multicolumn{4}{l}{\textbf{hLDA}} \\
\midrule
InfoVAST & 0.218 & 0.826 & 0.811 \\
Neurips  & 0.069 &	0.071 &	0.743 \\
BBC & 0.188 & 0.553  & 0.748\\
20News & 0.31 & 0.49 & 0.52 \\
Enron & 0.081 & 0.394 & 0.858 \\
Amazon  & 0.065 & 0.154 & 0.935\\
WOS46985  & 0.091 & 0.499 & 0.779\\
AGNews  & 0.149 & 0.331 & 0.921\\
\midrule
\multicolumn{4}{l}{\textbf{TSNTM}} \\
\midrule
InfoVAST & 0.08 & 0.19 & 0.28\\
Neurips  & 0.91 & 0.17 & 0.12 \\
BBC & 0.26 & 0.32 & 0.3 \\
20News & 0.31 & 0.49 & 0.52 \\
Enron & 0.18 & 0.29 & 0.38 \\
Amazon  & 0.20 & 0.38 & 0.38\\
WOS46985  & 0.19 & 0.37 & 0.31\\
AGNews  & 0.22 & 0.50 & 0.67\\
\midrule
\multicolumn{4}{l}{\textbf{hARTM}} \\
\midrule
InfoVAST & 0.15 & 0.59 & 0.72\\
Neurips & 0.23 & 0.32 & 0.67\\
BBC & 0.36 & 0.58 & 0.73\\
20News & 0.49 & 0.83 & 0.95\\
Enron & 0.40 & 0.72 & 0.85\\
Amazon  & 0.53 & 0.88 & 0.96\\
WOS46985 & 0.42 & 0.81 & 0.96\\
AGNews  & 0.52 & 0.87 & 0.95\\
\bottomrule
\end{tabular}
\end{adjustbox}
\caption{Topic Specialisation for other models}
\label{tab:topic_special_others}
\end{table}

\begin{figure*}[h!]
\includegraphics[width=\textwidth, scale=5]{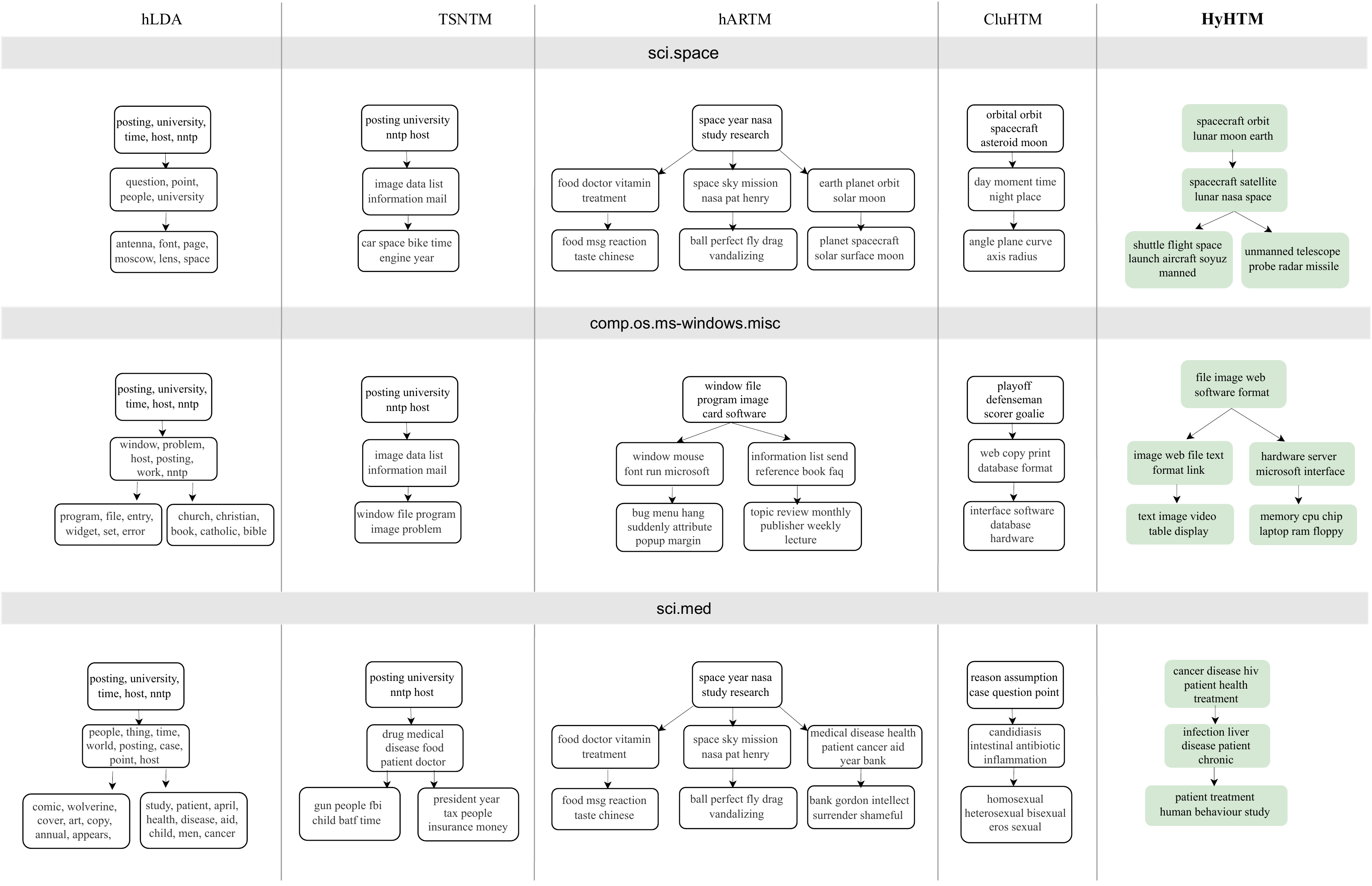} 
\caption{Comparing topic hierarchies for 20News documents. Every topic is represented by the top most probable words of the topic.}
\label{fig:space_hierarchy}
\end{figure*}

\section{Additional Ablation Study}\label{sect:ablation}

\noindent \textbf{Hierarchical clustering with Hyperbolic Embeddings:} 

\noindent We replace the underlying topic model algorithm with BERTopic \cite{grootendorst2022bertopic} which uses an HDBSCAN hierarchical clustering method under the hood which does not take into account the hierarchy between words in higher-level topics and lower-level topics. Both our model and BERTopic employ hyperbolic document embeddings as $A_0$, followed by their respective approaches to generate a hierarchy of topics. As seen in Table \ref{table:ablation3}, our model outperforms BERTopic in terms of coherence and hierarchical coherence measures. While the lower-level topics in BERTopic are related to their higher-level topics, the topic pairs (parent, child) were not unique as compared to our model.

\begin{table}[h!]
\centering
\begin{adjustbox}{max width=0.5\textwidth} 
\begin{tabular}{lccc}
\toprule
& { \centering HyHTM }  & \multicolumn{1}{p{1.5cm}}{ \centering HyHTM \\ c-TFIDF } & { \centering BERTopic }  \\
\midrule
{Coherence} & 0.325 & 0.269 & 0.293 \\
{Hierarchical Coherence} & 0.296 & 0.148 & 0.239 \\
\bottomrule
\end{tabular}
\end{adjustbox}
\caption{Ablation Study analyzing the effectiveness of our approach using the 20News dataset.}
\label{table:ablation3}
\end{table}

\noindent \textbf{Investigating the Need for Post-Processing Techniques in HyHTM for Ensuring Uniqueness Across Topic Levels: } 

\noindent BERTopic \cite{grootendorst2022bertopic} employs a class-based TFIDF approach for topic-word representation, treating all documents in a cluster as one. Inspired by this, we examined the impact of applying a similar class-based TFIDF to topics generated by our model as an additional post-processing step.  Theoretically, this should ensure unique topics at each level. However, as reported in Table \ref{table:ablation3} under \textbf{HyHTM c-TFIDF}, we found no noticeable improvement in topic coherence and hierarchy. This affirms that HyHTM inherently organizes documents into diverse and coherent themes at every level, obviating the need for additional post-processing.

\section{Implementation Details} \label{sect:implementation_details}

\subsection{Preprocessing}\label{sect:preprocessing}
We remove numeric tokens, punctuations, non-ascii codes and convert the document tokens to lowercase. In addition to NLTK’s stopwords, we also remove smart stopwords \footnote{\href{https://github.com/andersjo/pyrouge/blob/master/tools/ROUGE-1.5.5/data/smart_common_words.txt1}{Smart stopwords}} Next we lemmatise each token using NLTK’s WordNetLemmatizer. We filter the vocabulary  by removing tokens whose ratio of total occurrence count to number of training documents in which the token appears is less than 0.8. 

\subsection{Computing Infrastructure}
The experiments were run on a machine with NVIDIA GeForce RTX 3090 GPU and 24 GB of G6X memory. However, these experiments can also be replicated on CPU. The CUDA version used is 11.4.

\subsection{HyHTM} 

All experiments were performed with three runs per dataset. We use the implementation provided by \cite{stravzar2016orthogonal} for NMF. With this implementation we can leverage GPUs which helps us speed the topic model. \citet{viegas2020cluhtm}'s implementation utilises the scikit-learn \cite{scikit-learn} implementation of NMF. We report the difference in speed for both the approaches in Experiments \ref{sect:speed}.

\subsubsection{Varying \textbf{$k_H$: }Neighbourhood of a word defined in the hierarchical matrix} \label{sect:kHierarchy}

\begin{figure}[h!]
\includegraphics[width=0.5\textwidth]{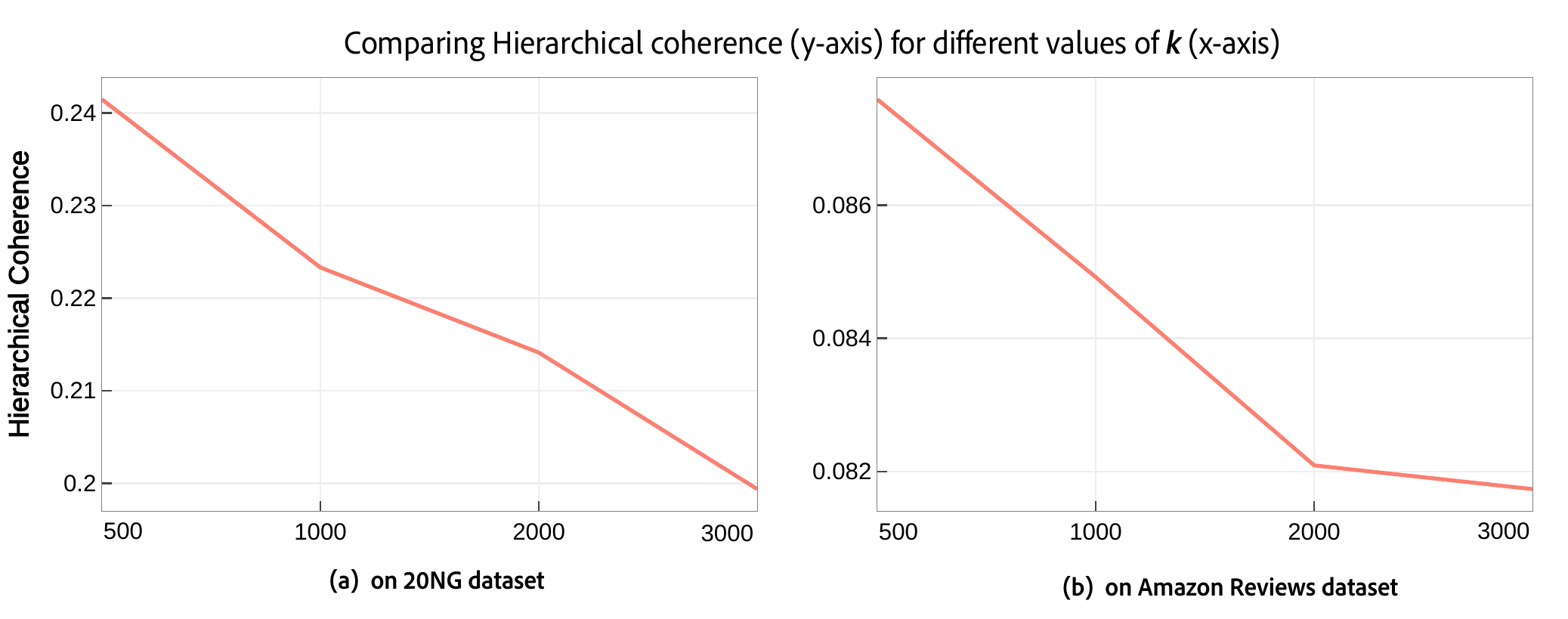}
\caption{$K_H$=500 performs the best out of all the choices on Hierarchical Coherence. A similar trend is observed on other metrics as well}
\label{fig:k-ablation}
\end{figure}

The term $k_H$ in equation \eqref{hierarchy} defines a neighborhood around words which helps us extract concept and sub-concept relations from hyperbolic geometry. If very large values of $k_H$ are considered, every word would be in the neighborhood of every other word, and for very small values of $k_H$, even though some very similar words will be included in the neighborhood, the overall document representation will become very sparse, and many concept and sub-concept relations are discarded.  We empirically tested this for $k_H$ in the range [500, 3000], and show our findings in figure \ref{fig:k-ablation}. We observe that when $k_H$ is 500, the hierarchical coherence along with the other metrics, is the highest, and after that, it drops.

\subsubsection{Varying \textbf{$\alpha$: } Similarity threshold in the similarity matrix} \label{sect:alpha}

The similarity threshold $\alpha$ in equation \eqref{similarity} is a hyperparameter that controls the pairs of words that should be considered similar and used to create the document representation. When the value is very high, only the most similar words are included in the term similarity matrix, which will result in a very sparse matrix, and defeat the purpose of adding more context about words from pretrained embeddings. If the value is very low, words which are not very similar can be picked up by the topic models as similar words. It is also important to note that while the vocabulary of terms can be controlled depending on the corpus used for topic modeling, the embeddings are pre-trained on large corpora which can result in biases from these corpora seeping into the arrangements of words in the embedding space. 

\begin{figure}[h!]
\includegraphics[width=0.5\textwidth]{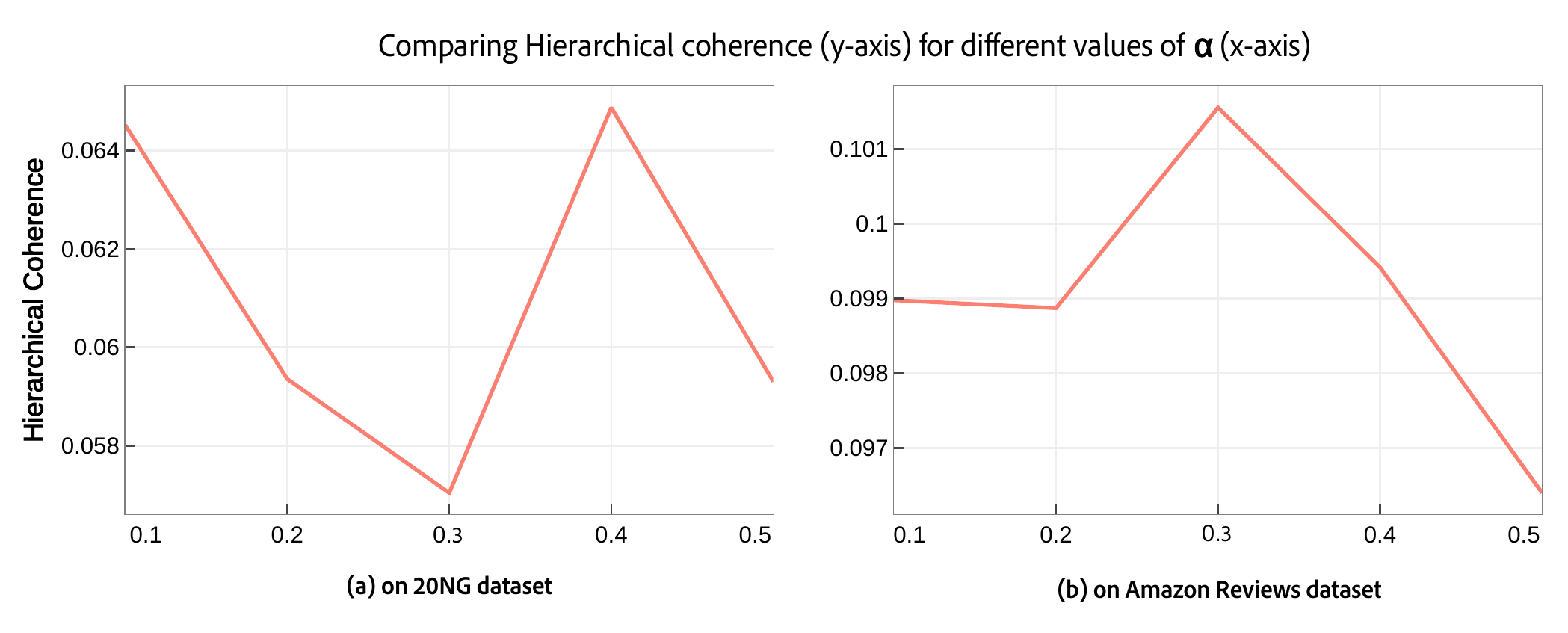}
\caption{$\alpha$=0.4 performs the best out of all the choices on Hierarchical Coherence. A similar trend is observed on other metrics as well}
\label{fig:alpha-ablation}
\end{figure}

We test our model with values of $\alpha$ that range from 0.1 to 0.5. In figure \ref{fig:alpha-ablation}, we observe that $\alpha$ value 0.4 gives the maximum value of hierarchical coherence for 20ng, and $\alpha$ value 0.3 is the maximum for Amazon Reviews. Similarly, we fine-tuned for all other datasets and report the results in the table \ref{tab:hyperparameter}.

\begin{table}[h!]
\centering
\begin{adjustbox}{max width=0.5\textwidth} 
\begin{tabular}{@{}llll@{}}
\toprule
\textbf{Dataset} & \textbf{$\alpha$} & \textbf{$k_h$} & \textbf{$k_S$}\\
\midrule
InfoVAST & 0.4 & 100 & 1000 \\
Neurips  & 0.4 & 100 & 500 \\
BBC & 0.4 & 100 & 500 \\
20News & 0.1 & 500 & 500 \\
Enron & 0.4 & 100 & 500 \\
Amazon  & 0.3 & 500 & 500\\
WOS46985  & 0.1 & 100 & 500\\
AGNews  & 0.1 & 500 & 500\\
\bottomrule
\end{tabular}
\end{adjustbox}
\caption{Best performing hyperparameters.}
\label{tab:hyperparameter}
\end{table}

\subsection{CluHTM}
We use the implementation provided by \cite{viegas2020cluhtm}\footnote{\href{https://github.com/feliperviegas/cluhtm}{https://github.com/feliperviegas/cluhtm}} for the CLUHTM baseline. While this implementation does provide a method to learn the optimal number of topics, it is highly inefficient, taking $\mathcal{O}(n^3)$ time. The training time for this model on 20NG data was $\approx$ 32 hours, and AR was $\approx$ 22 hours. Additionally, the number of topics is different in every branch, and comparison across models becomes difficult. 

\subsection{hARTM}
For the hARTM baseline model, we use the BigARTM\footnote{\href{https://bigartm.readthedocs.io/en/stable/}{BigARTM}} package, version 0.10.1. For this model, we cannot choose the number of subtopics explored for each parent, but we can control the total number of subtopics from all parents at a certain level. In our other parametric models, since each parent has $n$ subtopics, we obtain a total of $n^l$ topics at level $l$. Thus for hARTM, we indicate that the model chooses $n^l$ topics at level $l$ starting from $l=1$ to a depth of $l=3$.

\subsection{hLDA}
We use the following implementation\footnote{\href{https://github.com/joewandy/hlda}{hLDA codebase}} for hLDA. 

\subsection{TSNTM}
We use the official implementation provided by \cite{isonuma2020tree} \footnote{\href{https://github.com/misonuma/tsntm}{TSNTM codebase}} for TSNTM.

\subsection{BERTopic}
We use the official implementation provided by \cite{grootendorst2022bertopic} \footnote{\href{https://github.com/MaartenGr/BERTopic}{BERTopic codebase}} for BERTopic. We use the default parameters setup by BERTopic for HDBSCAN clustering. 

\section{Number of topics for parametric models}\label{sect:numberOfTopics_parametric}

For the parametric models like hARTM, CluHTM, and our model HyHTM, we use the same number of topics at every level for a fair comparison. We explain how the topic hierarchy grows when the number of topics at each node of the tree is $N = 10$. 

\begin{enumerate}
    \item At the root level (level 1), we train the model on the entire corpus of documents $D$ and set the number of topics as $N=10$. As a result, we get 10 topics at the root level. 
    
    \item For every topic in the previous level, each parametric model organizes how documents will get distributed across topics. For CluHTM and HyHTM, a document is assigned to the topic with which it has the maximum association. Therefore, each document is assigned only 1 topic at a given level. Once the documents are categorized, we perform NMF on these documents and produce 10 topics for every parent topic. 
\end{enumerate}
In this way, we obtain topics at root level as 10, level 2 as $10^2=100$, and level 3 as $10^3=1000$.  hARTM follows a different procedure using regularisers for categorizing documents and exploring lower-level topics. After level 1, hARTM produces flat topics in level 2 and learns the association between every lower-level topic with the higher-level topic. We assign the number of topics in level 2 as $10^2$, the same as the total number of topics in level 2 for CluHTM and HyHTM, and similarity for level 3.

\section{Number of topics for Non-Parametric models}\label{sect:numberOfTopics_nonparametric}

The number of topics for non-parametric models is listed in Table \ref{non-param-topics}:

\begin{table}[h!]
\centering
\begin{adjustbox}{max width=0.5\textwidth} 
\begin{tabular}{@{}llllll@{}}
\toprule
\textbf{Dataset} & \textbf{Model} & \textbf{Total topics} & \textbf{L1 topics} & \textbf{L2 Topics} & \textbf{L3 topics} \\ \midrule
InfoVAST          & hLDA           &              15         &          1         &          4          &         10           \\
                 & TSNTM          &      12 & 1 & 5 & 6 \\ \midrule
Neurips             & hLDA           &   6 & 1 & 1 & 4 \\
                 & TSNTM          &     14 & 1 & 4 & 9 \\ \midrule
BBC          & hLDA           &      35 & 1 & 7 & 27 \\
                 & TSNTM          &      8 & 1 & 3 & 4 \\  \midrule
20News             & hLDA           &     122 & 1 & 14 & 107 \\
                 & TSNTM          &     20 & 1 & 7 & 12 \\
\midrule
Enron            & hLDA           &   194 & 1 & 15 & 178 \\
                 & TSNTM          &     9 & 1 & 3 & 5 \\\midrule
Amazon       & hLDA           &      395 & 1 & 16 & 378 \\ 
                 & TSNTM          &    11 & 1 & 4 & 6 \\\midrule
WOS          & hLDA           &   38 & 1 & 8 & 29 \\
                 & TSNTM          &      11 & 1 & 4 & 6 \\ \midrule
AGNews           & hLDA           &    344 & 1 & 16 & 327 \\
                 & TSNTM          &   14 & 1 & 5 & 8 \\ 
\bottomrule
\end{tabular}
\end{adjustbox}
\caption{Number of topics for non-parametric models}
\label{non-param-topics}
\end{table}

\end{document}